\documentclass[12pt,preprint]{aastex}

\slugcomment{Not to appear in Nonlearned J., 45.}
\slugcomment{\ \ \ \ \ \ \ \ \ \ \ \ \ \ \ \ \ \ \ \ \ To appear in ApJS v165n1, July 2006.\\
{\tiny{-}}\ \ \ \ \ \ \ \ \ \ \ \ \ \ \ \ \ \ \ \copyright American Astronomical Society, 2006.}

\shorttitle{Comparing Pattern Recognition Feature Sets}
\shortauthors{Proctor}

\begin{document}




\title{Comparing Pattern Recognition Feature Sets for Sorting Triples in the FIRST Database}


\author{D. D. Proctor}
\affil{Institute of Geophysics and Planetary Physics, L-413, Lawrence Livermore National Laboratory, \\
7000 East Avenue, Livermore, CA, 94550; proctor1@llnl.gov}


\begin{abstract}
Pattern recognition techniques have been used with increasing success for coping with the tremendous amounts of data being
generated by automated surveys.  Usually this process involves construction of training sets, the typical examples of data with known classifications.  Given a feature set, along with the training set, statistical methods can be employed to generate a classifier.  The classifier is then applied to process the remaining data.  Feature set selection, however, is still an issue.  This report presents techniques developed for accommodating data for which a substantive portion of the training set cannot be classified unambiguously, a typical case for low resolution data.  Significance tests on the sort-ordered, sample-size normalized vote distribution of an ensemble of decision trees is introduced as a method of evaluating relative quality of feature sets.  The technique is applied to comparing feature sets for sorting a particular radio galaxy morphology, bent-doubles, from the Faint Images of the Radio Sky at Twenty Centimeters (FIRST) database.  Also examined are alternative functional forms for feature sets.  Associated standard deviations provide the means to evaluate the effect of the number of folds, the number of classifiers per fold, and the sample size on the resulting classifications.  The technique also may be applied to situations for which, though accurate classifications are available, the feature set is clearly inadequate, but is desired nonetheless to make the best of available information.
\end{abstract}





\section{INTRODUCTION}

Successful development of pattern recognition classifiers for use on the the vast amount of
data generated by automated surveys is being increasingly reported.
Recent examples are \citet{RG} for spectral classification, \citet{BA} and \citet{DBM} for morphological galaxy classification, \citet{JC} for extended source classification and \citet{CMHF} and \citet{OCG} for star/galaxy discrimination.
While some discussion of feature set selection has been made in these and other papers, feature set selection continues as a topic of research interest in pattern recognition procedure.
\citet{BDMT} have stated that the precise choice of features is perhaps the most difficult task in pattern processing.  
Regarding the choice of merit function, \citet{BJain}, in their comprehensive review of statistical pattern recognition, state that most feature selection methods use the classification error of a feature subset to evaluate its effectiveness.
However, for applications in which only a portion of the training set can be accurately classified (low resolution applications, for example) recognition rates and classification errors are problematical and other approaches are necessary.  In the author's initial paper (Proctor 2002; hereafter Paper I) on low resolution pattern recognition, five, nine, fifteen and twenty-one member features sets were compared, using recognition rates, for the sorting of a particular radio galaxy morphology. 
As part of a process of looking for intrinsic characteristics of the target class, it is of interest to eliminate extraneous features, the concern being extraneous features will degrade the solutions.

While detailed discussion is given below, at this point some definitions are in order.  Briefly, statistical 
methods are applied to generate a classifier using typical data of known classification. 
Application of the classifier to a sample member of unknown classification results in the member being assigned an estimate of the probability of its being a particular class.  When multiple classifiers are generated using some randomization procedure, the average of the resulting probability estimates for the member will be designated the normalized score or vote.  Ordering this score, say high to low, for the entire sample, results in a sort-ordered distribution, with index 1 to N, where N is the number of members in the sample.  If the index is divided by the number of sample members it becomes the sort-ordered index, sample size normalized.  It is the plot of normalized score or vote versus the sample-size-normalized index that is designated the vote curve.

In Paper I, vote curves were used to examine the ability of the decision tree classifier to generalize to previously unseen samples.  This was accomplished by comparing this distribution for the training set with that of the test set.  In this report, vote curves are used to compare feature sets of particular interest in the application.  
The focus will be on comparison of feature sets using multiple runs of Oblique Classifier One (OC1), the decision tree software of \citet{BMKS}.  Besides being freely available, \citet{BWhite} found oblique decision trees represented a good compromise between demands of computational efficiency, classification accuracy, and analytical value of results.  \citet{BSCFMW} report OC1 classification accuracy comparable to CART \citep{BBreiman} and C4.5 \citep{BQuinlan} for automated identification of cosmic-ray hits in Hubble Space Telescope images.  It has also been adopted for the 2Mass extended source catalog \citep{JC}.
More general discussions of the feature selection and evaluation process can be found in \citet{BJZ}, \citet{BCVC} and \citet{BNF}.

This report is organized as follows:  The background of the pattern recognition application and a summary of Paper I are presented in Section 2.  Section 3 describes a series of feature set comparisons using the vote curves.  Finally, Section 4 contains discussion and conclusions.

\section{BACKGROUND OF PATTERN \\
RECOGNITION CASE STUDY}

\begin{figure*}[h]
 \centering
 \plotone{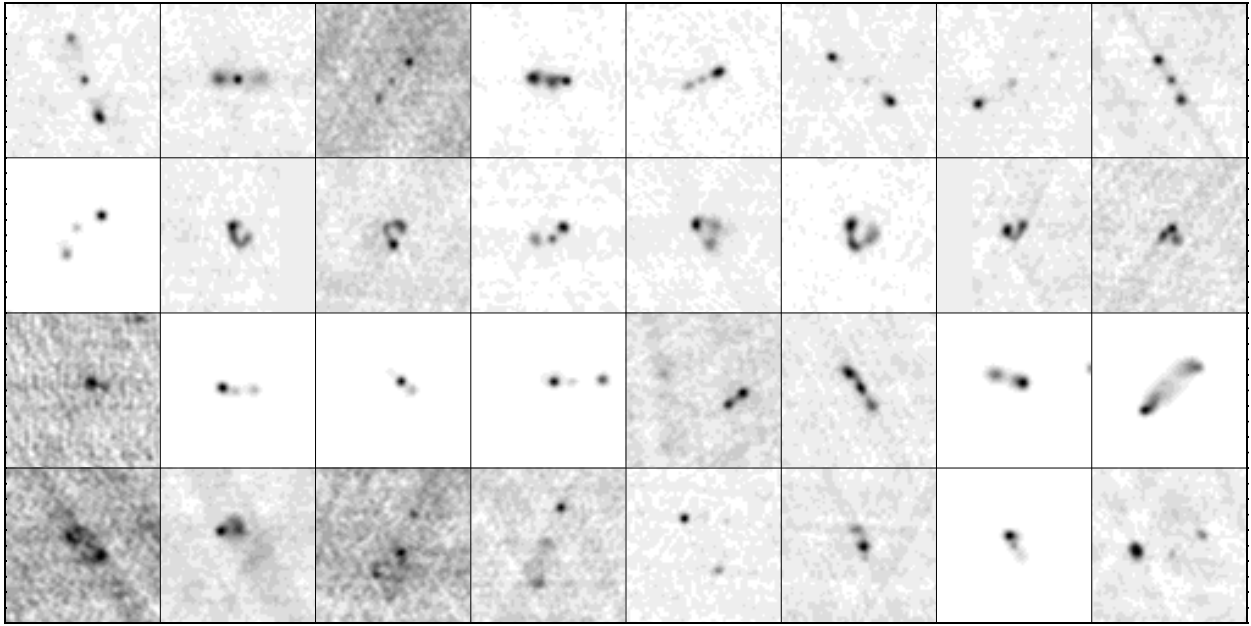}
 \caption{\label{fig:epsartcc} Class examples: Top row, prototypical three component radio galaxies.  Second row, bent three component radio galaxies.  Third row, nonbent three component sources.  Last row, ambiguous sources.} \label{Fcc}
\end{figure*}

The pattern recognition problem under consideration is the selection of a particular type of three component radio galaxy, the so called "bent doubles".  It is believed these bent radio galaxies can act as tracers of rich clusters and clusters at high redshift \citep{Bl}.  A proto-typical three component radio galaxy consists of two jets or lobes extending from opposite sides of a central core.  Examples are shown in the first row of Figure~\ref{Fcc}.  For bent doubles the jets or lobes appear swept back as by a wind.  The second row of Figure~\ref{Fcc} shows examples of this target class.  The target class is to be separated from nonbent, S-shaped, and chance-projection three component sources.  Examples of nonbent three component sources are shown in the third row of Figure~\ref{Fcc}.  The final row of the figure shows examples of ambiguous sources, those for which visual classification is uncertain due to poor resolution or low signal-to-noise ratio.  The data used in this study come from the images and catalog \citep{BWBHG} developed by the Faint Images of the Radio Sky at Twenty Centimeters (FIRST) Survey \citep{BBWH} collaboration.  The catalog includes source position, fitted parameters relating to source size and flux density, and noise estimates for each component.  Sample entries for the three components of the first image in Figure~\ref{Fcc} are given in Table 1.

\begin{table*}[h]
\begin{center}
\caption{\label{tab:table1} Catalog Entries for first image, top row of Figure 1}
\end{center}
\begin{tabular}{ccrrrrrrc}
\tableline\tableline
RA(2000)\tablenotemark{a}&Dec(2000)\tablenotemark{b}& Sp\tablenotemark{c}\  &Si\tablenotemark{d}\  &\ RMS\tablenotemark{e}\ &Maj\tablenotemark{f}&Min\tablenotemark{g}&PA\tablenotemark{h}\ &Field\tablenotemark{i}  \\
\tableline
07 14 52.780& +37 43 17.29&  5.44& 11.15& 0.135& 9.67&  6.18& 23.7& 07150+37367F\\
07 14 53.694& +37 43 44.21&  4.16&  4.25& 0.135& 5.80&  5.12& 11.7& 07150+37367F\\
07 14 54.505& +37 44 14.00&  2.69&  5.75& 0.136& 9.12&  6.82& 20.5& 07150+37367F\\
\tableline
\end{tabular}
\tablenotetext{a}{Right Ascension (hr min sec).}
\tablenotetext{b}{Declination (deg min sec).}
\tablenotetext{c}{Peak flux density (mJy/beam)}
\tablenotetext{d}{Integrated flux density (mJy) for Gaussian fit to source.}
\tablenotetext{e}{Estimated RMS noise (mJy/beam) at source position.}
\tablenotetext{f}{Fitted major axis FWHM (arcsec).}
\tablenotetext{g}{Fitted minor axis FWHM (arcsec).}
\tablenotetext{h}{Position angle of major axis (degrees east of north).}
\tablenotetext{i}{Name of image including this source.}
\end{table*}

  A random sample of 2823 sources were selected from the available population of about 15,000
three-component sources.  Each source was visually assigned to bent double, nonbent double or the ambiguous class, the counts being $N_{bent}$=147, $N_{nonbent}$=1395, and $N_{amb}$=1281 respectively.  This sample is designated the training/test set.  The training set consists of only the visual bent and nonbent sources exceeding signal-to-noise ratio of 8.5, composed of $N_{bent,tr}$=115 visual bents and $N_{nonbent,tr}$=930 visual nonbents and excludes ambiguous sources.  (The signal-to-noise ratio is defined as the peak flux of the component having the smallest peak flux divided by its root-mean-square error.)  That a significant portion of the training/test set was assigned an ambiguous classification was attributed to the relatively low resolution (in pattern recognition terms) of the survey, 99 percent of components having fitted major axis less than 12 pixels.  It was felt that since ambiguous sources are such a large fraction of the sample, rather than guess on their visual classification, training on the more reliable sources would improve signal-to-noise for the classifier.  Subsequent comparison of vote distributions determine the extent to which this is justified.  An alternative approach might be to modify decision tree construction to include weighting of training set classifications. Since the complexity of this alternative was unclear, examination of this approach was deferred.  While a three class (bent, nonbent, ambiguous) classifier could have been constructed the two-class approach serves to force ambiguous sources into bent or nonbent classes, more in correspondence with the physical situation.  

A comment on the size of the training/test set sample is due.  Given the complex interrelationship between sample size, number and characteristics of features and classifier complexity, guidance for sample size is difficult to establish apriori.  \citet{BJainC} recommend using at least ten times as many training samples per class as the number of features,  with larger ratios for more complex classifiers.  For this problem with low resolution, lack of scale and orientation information, chance superposition of sources, and considerable variation in bent morphology it was felt the more the better.  The sample is a result of an approximately eight hour day spent classifying a random selection of three component sources.  This allowed for spending an average of about ten seconds per image.

One of the classifiers studied in Paper I was
Oblique Classifier One (OC1).  It is a system to generate a decision tree from a training set of numerical features of known classes, attempting to produce a tree that has pure samples of training set objects.  OC1's default impurity measure, the twoing rule \citep{BBreiman}, was used. (The impurity measure is the metric that is used to determine the "goodness" of a hyperplane location.)

\begin{table*}[!h]
\begin{center}
\caption{List of Features for five, nine, fifteen and twenty-one member feature sets}
\end{center}
\label{feature_list}
\centering
\begin{tabular}{rll}
\tableline\tableline
No.& Description \\
\tableline
1. & $d_{mid}$ .......................... intermediate length of pairwise distances between components. \\
2. & $d_{min}$/$d_{mid}$ ................ ratio of smallest distance to intermediate distance. \\
3. & ($d_{mid}$+$d_{min}$)/$d_{max}$ .... ratio of sum of intermediate and smallest distances to largest distance. \\
4. & $R_{ss}$............................ ratio of silhouette sizes of assumed lobes or jets (smaller to larger). \\
5. & $T_{ss}$ ........................... total calculated silhouette area, all three components. \\
6. & Ratio distance between midpoint of shortest leg and its opposite source to length of shortest leg.\\
7. & Absolute value of cosine of angle between major axis and direction of core for source opposite \\
   & intermediate leg.\\
8. & Ratio of square root of silhouette size to length of opposite leg for core.\\
9. & Ratio of square root of silhouette size to length of opposite leg for source opposite \\
   &intermediate leg.\\
10. & Silhouette size of arm with maximum silhouette size.\\
11. & Absolute value of cosine of angle between the major axis of the source opposite the smaller \\
    & arm and the direction of the presumed core.\\
12. & Integrated flux of the core.\\
13. & Integrated flux of source with maximum integrated flux.\\
14. & Integrated flux of source opposite longer arm.\\
15. & Ratio of integrated flux of minimum integrated flux of the arms to maximum of integrated \\
    & flux of the arms.\\
16. & Ratio peak flux to integrated flux of secondary.\\
17. & Ratio peak flux to integrated flux of weakest source.\\
18. & Ratio integrated flux of primary to total flux of three components.\\
19. & Ratio integrated flux of weakest source to total flux of three components.\\
20. & Ratio integrated flux of core to total flux of three components.\\
21. & Ratio integrated flux of source opposite shortest arm to total flux.\\
\tableline

\end{tabular}
\end{table*}

\begin{figure}[h]
 \centering
 \includegraphics{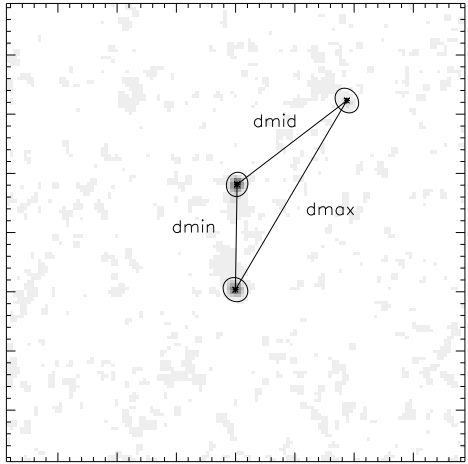}
 \caption{\label{fig:epsartq} Projected geometry, three component source. The core is assumed opposite side of size dmax, the other sources are presumed lobes or jets or chance projections.} \label{Fq}
\end{figure}

An initial set of five basic features was used to generate classifiers and subsequently features were added.  Five, nine, fifteen and twenty-one member feature sets were used.  The added features were mostly more contrived expressions for which a visual examination of distributions for bent and nonbent sources appeared different.  The features used were all derived from catalog entries of the three components.  Table 2 gives the features used.  Distances are projected distances on the plane of the sky.  The geometry is illustrated in Figure~\ref{Fq}.  The core is assumed to be the component opposite the longest leg of the triangle formed by the three components, the other components being possible lobes or jets.  The silhouette sizes are calculated by evaluating the number of pixels with flux density greater than the threshold for a model calculated from the catalog entries of the component.  The fitted model functional form of the flux density S(x,y) at position (x,y) is given by
\begin{eqnarray}{\rm  S(x,y)=Sp }\exp (-({x^2 \over 2\sigma _x^2} + {y^2 \over 2\sigma _y^2})),\end{eqnarray}
where Sp, ${\sigma _x}$ and ${\sigma _y}$ are derived from catalog entries for the component.   The number of pixels greater than a threshold is then calculated to determine the silhouette size of the component and the appropriate ratio taken for  $R_{ss}$, the ratio of silhouette sizes, smaller to larger.

Cross validation was used, with the training/test set being divided into five folds.  The training set members from four folds were used to classify the entire remaining fold, each fold thus being classified in succession from the classifiers generated by the other four folds.  Cross validation is a standard pattern recognition technique used to avoid bias that would be introduced if the points used in testing were the same as those used for training.
The OC1 search algorithm includes some randomization to avoid local extrema in the search space.  
In generating the tree, OC1 first selects a random initial location at each node and adjusts until a locally optimal hyperplane is reached.  Multiple random starting points are used at each node and the hyperplane is perturbed in random directions after it reaches local minimum.

Heath, Kasif, and Salzberg \citep{BHKS} have shown the accuracy of classification is improved by having multiple trees vote.  Thus, ten classifiers were generated, using different seeds for the random number generator, for each of the five folds, resulting in a total of fifty decision trees for each feature set.

Typically decision trees are pruned to avoid overfitting of the data.  With OC1, a randomly chosen subset of the training set is reserved for use in pruning the tree.  For this study the pruning portion was 20\%.  Details of OC1 pruning can be found in \citet{BMKS}.


When the five, nine, fifteen and twenty-one feature classifiers were compared, recognition rates and false positives were within about one mean square error of each other using a somewhat arbitrary top 16\% of the vote curve being classified bent.  In the following section we present more extensive comparisons of vote curves for these feature sets and others.
Only the issue of comparing feature sets with OC1 decision tree vote distributions will be addressed, though the technique should be applicable to any method employing randomization in the construction of the classifier.

\section{SOME FEATURE SET COMPARISONS}

An estimate of the probability in favor of the object being of the bent class could be made by generating numerous trees and dividing the number of times the object was classified as the bent by the number of trees.  For this study, each tree's vote on a source was apportioned according to the prescription followed by \citet{BWBG} for pruned decision trees.  Using this prescription, if a sample ends up at a leaf node with $N_{L}$ training set objects of which B are bent, the tree's single vote on the source is split into the fraction (B+1)/($N_{L}$+2) in favor and the fraction ($N_{L}$-B+1)/($N_{L}$+2) against bent classification.  This form was also adopted by \cite{BMcG}.  They note the ratio is derived from the binomial statistics at the leaf.  The votes of the ten trees in favor of a source being bent were then averaged.  It is this normalized score or vote, shown in subsequent comparison plots, that provides an estimate of the probability of individual three-component source being of the bent class.  It is emphasized this is a conditional probability, depending on classifier, feature set, and training set.

   For each source i, if $P_{i}$(bent) is the estimate in favor of
    the source being bent and $P_{i}$(nonbent) is the estimate against,

             $P_{i}$(bent) + $P_{i}$(nonbent) = 1.

    Thus 

             ${\Sigma_i}$ $P_{i}$(bent)  + ${\Sigma_i}$  $P_{i}$(nonbent)  = N,

     where N is the number of points in the sample under consideration.

     Normalizing by N gives

          ( ${\Sigma_i}$ $P_{i}$(bent) + ${\Sigma_i}$ $P_{i}$(nonbent) ) /N  =  1.

     This normalized total in favor is the
     'area under the curve' of the sort-ordered, sample-size normalized vote plot.

  For accurate classification and adequate features, it is expected the unpruned decision tree when acting on the training data would produce $N_{bent,tr}$ sources classified as bent and $N_{nonbent,tr}$ sources classified as nonbent. 
This leads to the expectation that, ideally, the area under the training set vote curve should be $N_{bent,tr}$/($N_{bent,tr}$ + $N_{nonbent,tr}$) and thus constant for a given training set.  Thus evaluation of feature sets can be based on the shape of the vote curve as well as the comparison of vote distributions for the visual bents.  Ideally, the vote curve would start at 1.0 and drop vertically to 0.0 at the true, but for this application unknown, bent fraction, while the vote distribution for the training set bents  would be uniformly 1.0.  It should be noted that this ideal may not be attainable due to lack of sufficiently distinguishing features to break the degeneracy. 
One training/test vote curve will be referred to as more compact than a second if it has higher probabilities for the lower index ranges (expected bents) and lower probabilities for the higher index ranges (expected nonbents).
Vote curve comparisons follow a brief discussion of generalization.

Generalization is the ability of a classifier to classify previously unseen samples.   Usually it is implicit that the unseen sample has the same distribution in feature space as those used in classifier construction.  Depending upon the application this may or may not be a reasonable assumption.  In Paper I, for this application, the assumption was examined by comparison of the training set vote curve with the entire training/test set vote curve (fifteen feature classifier).  This (training set - training/test set) comparison seems reasonable, since by using cross validation, a source being classified is not used in the construction of its classifier.  Here the comparison is made for the five feature classifier.   It is also of interest to include comparison with the vote curve for all nontraining set points.    

\begin{figure*}[h]
 \centering
 \includegraphics{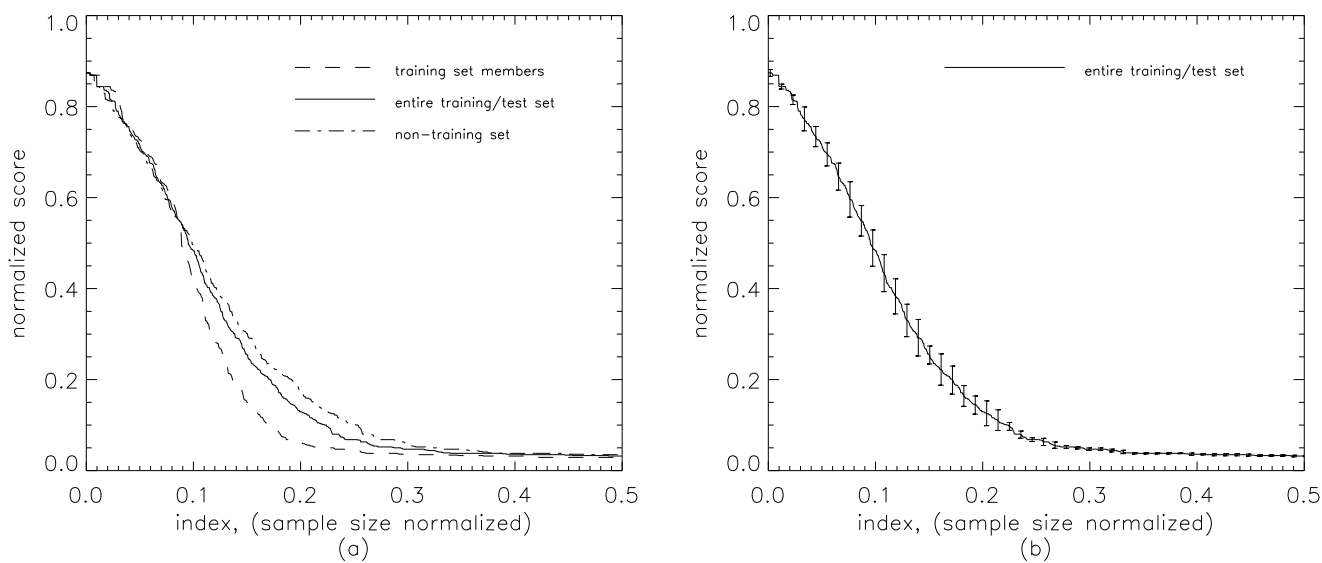}
\caption{\label{fig:epsarto} (a) Vote curve comparison of training set with entire training/test set, and the set that excludes points used in training (five feature classifier), (b) Error bars for selected points along entire training/test set vote curve.} \label{Fo}
\end{figure*}

Figure~\ref{Fo}a shows the comparison, for the five-feature classifier, of the training set vote curve, the training/test set vote curve, and the vote curve for all nontraining set points .
(Since the distributions were essentially flat after normalized index 0.5, only initial half of distribution is shown.) 
  Consistent with the fifteen feature comparison shown in Paper I, this five feature comparison suggests fairly good generalization from training to test set.
In Figure~\ref{Fo}a the area under the training sample curve is 0.112, compared with the bent fraction ($N_{bent,tr}$/($N_{bent,tr}$+$N_{nonbent,tr}$) = 0.110).  This compares with the area under the curve for the entire training/test set of 0.125 and the area under the distribution that excludes the training set points of 0.133.  These latter are the ambiguous population and those sample members with low signal-to-noise ratio.  The difference in area between the two extreme curves is approximately 17\%.  Note that the curves match well over approximately 80\% of the respective samples, from 0.0 to 0.1 and again from 0.3 to 1.0 normalized sort order index.  This seems consistent with what might be expected due to errors in the training set classification and resulting differences in bent fractions for the various samples.  A 10 to 20\% error in the visual classifications would not be that surprising.    A mismatch at the high or low ends of the vote curve, however, would be more suggestive of underlying differences between the populations.

Figure~\ref{Fo}b shows error bars at selected points along the vote curve.  The error bars are plus and minus one standard deviation of mean of the ten decision trees for the respective point's fold.  It thus represents the error associated with classifier construction for the given feature set and training set.  As might be expected the errors are smaller at the extremes.  In the range of normalized index 0.1 to 0.3 the maximum difference between training and nontraining set points is less than 0.15.

\subsection{\it Comparison of Five, Nine, Fifteen and Twenty-one Member Feature Sets}

\begin{figure*}[h]
 \centering
 \includegraphics{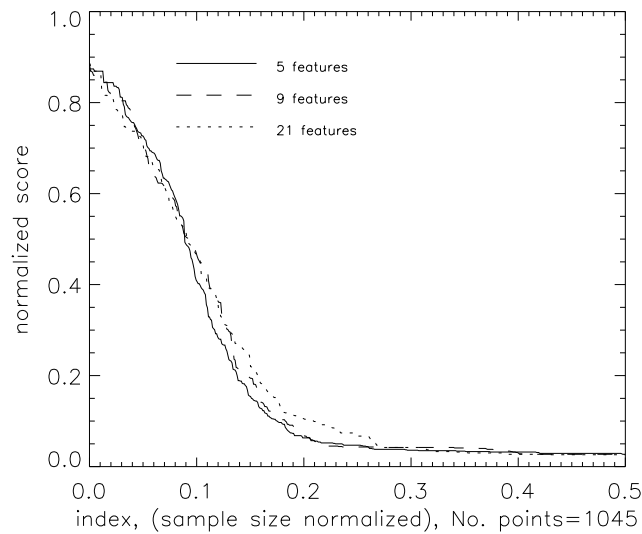}
 \caption{\label{fig:epsartaa} Vote curve comparison of five, nine, and twenty-one feature classifiers (training set). Fifteen feature vote curve intermediate between nine and twenty-one feature vote curve.} \label{Faa}
\end{figure*}

\begin{figure*}[h]
 \centering
 \includegraphics{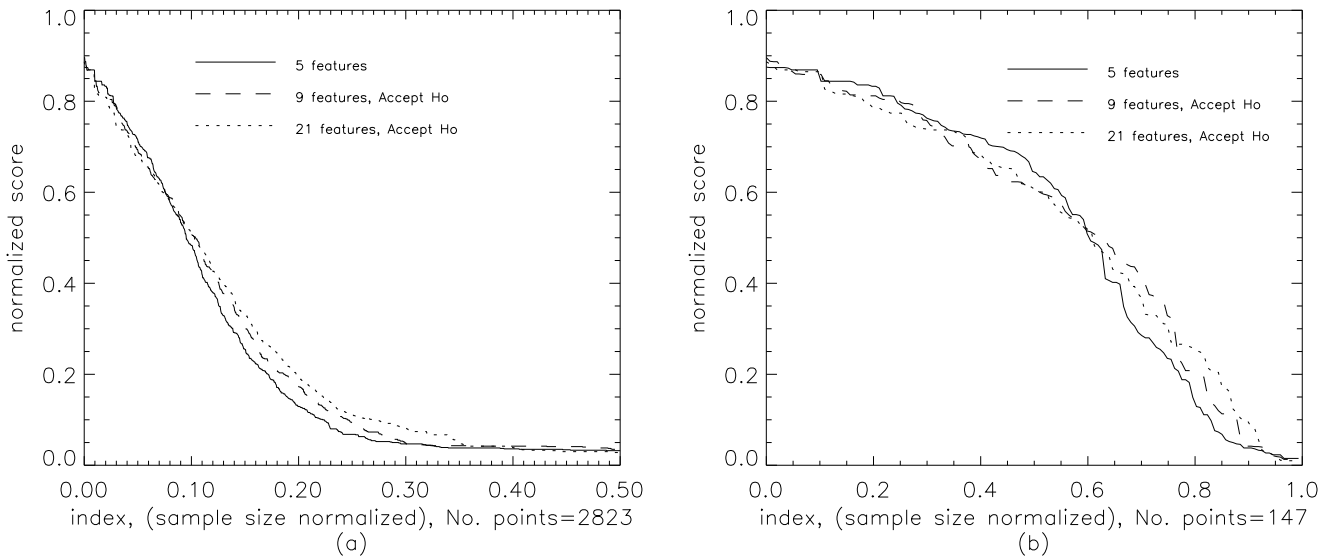}
 \caption{\label{fig:epsarta} Vote curve comparisons of nine, and twenty-one feature classifiers with five feature classifier, (a) entire training/test set (b) visual bents.} \label{Fa}
\end{figure*}

Figure~\ref{Faa} shows a comparison of the vote curves of the training set for five, nine, and twenty-one feature classifiers, whereas  Figure~\ref{Fa}a shows the curves for the entire
training/test set.  
While the distributions in Figure~\ref{Fa}a are broader, overall the relative order of the feature sets is the same for both figures.
The fifteen feature classifier distribution was intermediate between the nine and twenty-one feature classifier for both figures and was omitted to improve plot clarity.  
In this comparison, the five feature set distribution appears the most compact, and thus the most desirable.

In this and following comparisons, the results of statistical tests at the 5\% significance level are reported.  For the visual bent vote curve, as Figure~\ref{Fa}b, the Kolmogorov-Smirnov test, comparing two cumulative distribution functions, and the Wilcoxon signed rank test \citep{BOS}, comparing effects of two treatments on paired data, were applied.  If the statistical test results were in agreement, the mutual result is reported, if not, the results are listed in order Kolmogorov-Smirnov result and Wilcoxon signed rank result.  For the training/test set vote curves, as in Figure~\ref{Fa}a, score values above 0.05 are compared using Conover's distribution functions \citep{BCo} for Tsao's truncated Smirnov statistics \citep{BTs}.  The statistical tests are distribution-free tests.  They do not require the form of the distribution to be known.  Details and discussion of the selection of the statistical test for the training/test set are in Section 3.5.  
In all instances, the null hypothesis is $H_{o}$: no difference in distributions under consideration.

The significance tests show that, at the 5\% significance level, when each of the
nine, fifteen, and twenty-one feature training/test set vote curves is compared with the five
feature vote curve, Figure~\ref{Fa}a, the hypothesis of equivalent distributions is accepted.
Figure~\ref{Fa}b shows corresponding vote curves for visual bents only.  The significance tests show that at the 5\% significance level the hypothesis of equivalent distributions is accepted. 
These results appear consistent with noise introduced by inclusion of extraneous features causing slight degradation in the compactness of the vote distributions for the entire sample, but the classifier being able to generate substantially equivalent classifications for the visual bents.

\begin{figure}[h]
 \centering
 \includegraphics{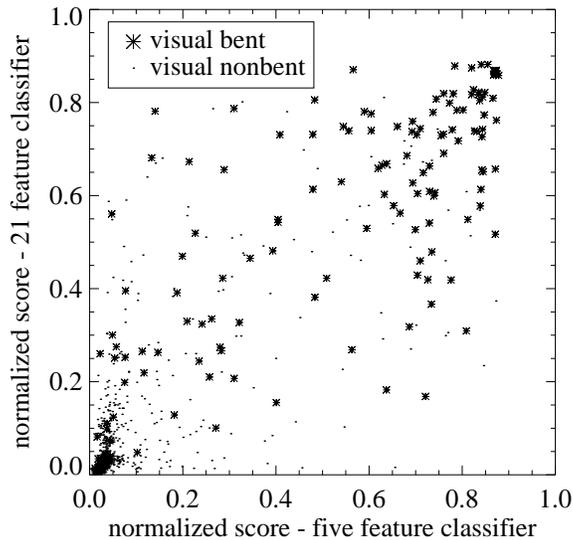}
 \caption{\label{fig:epsartc}Vote comparison of five and twenty-one feature classifiers.  Number of points=1542.  Eleven of 147 visual bent doubles had both classifier scores less than 0.05.  Most points are clustered in lower left corner.} \label{Fc}
\end{figure}

  Figure~\ref{Fc} is a direct comparison of
 the normalized score of the five feature classifier with the normalized score of the twenty-one 
feature classifier for each training point.  (A small normal random offset (sigma = .005) in the score was added to improve visualization, since at lower vote values many points overplot.)  While there is
relatively good agreement on most very low scoring sources (normalized vote less than 0.05 for both classifiers), there is considerable scatter in higher vote sources.
Correlation coefficients between the five and 21 feature classifier votes for the entire bent/nonbent training set is 0.88, whereas, 
for visual bents alone, the correlation coefficient is 0.79.  


For each of the subsequent comparisons, distributions for the training set showed the same relative order as the entire training/test set distributions.  Thus for subsequent comparisons, only the results of the entire training/test set will be shown.  

\subsection{\it Comparison of Five-Member Feature Set with its Various Four-Member Feature Subsets}
Since Figure~\ref{Faa} and Figure~\ref{Fa} suggest no substantial benefit from adding features to the original five member feature set, it is of interest to look at feature sets with fewer members.  
From a data-mining viewpoint, interest is in determination of intrinsic characteristics of the target class.  Interpretation of decision tree results is difficult with even as few as three features, since the number of decision trees per feature set is the number of folds times the number of trees per fold. 
Though resulting classifications may be similar, interpretation of results is simpler without extraneous features.


In general, given a feature set of size d, the feature selection problem is to determine the subset of d that produces the best classifier.  There are $ _{d}$$C_{m}$ = d!/(m!(d-m)!) possible subsets of size m.  The number of subsets grows combinatorially and exhaustive search becomes impractical for feature sets as small as seven or eight for this application, given current computer speeds.  Even for d=5 this amounts to 31 possible subsets ranging in size from m=1 to m=5.  Realizing that no nonexhaustive sequential feature selection procedure can be guaranteed to produce the optimal subset \citep{BCVC}, in lieu of examination of 31 feature subsets, a few comparisons will be explored.  Sequential backward selection will be applied to the five member feature set.  Sequential backward selection starts with the five features and successively deletes one feature at a time.  \cite{BJZ} discuss other well-known feature selection methods.

\begin{figure*}[h]
 \centering
 \includegraphics{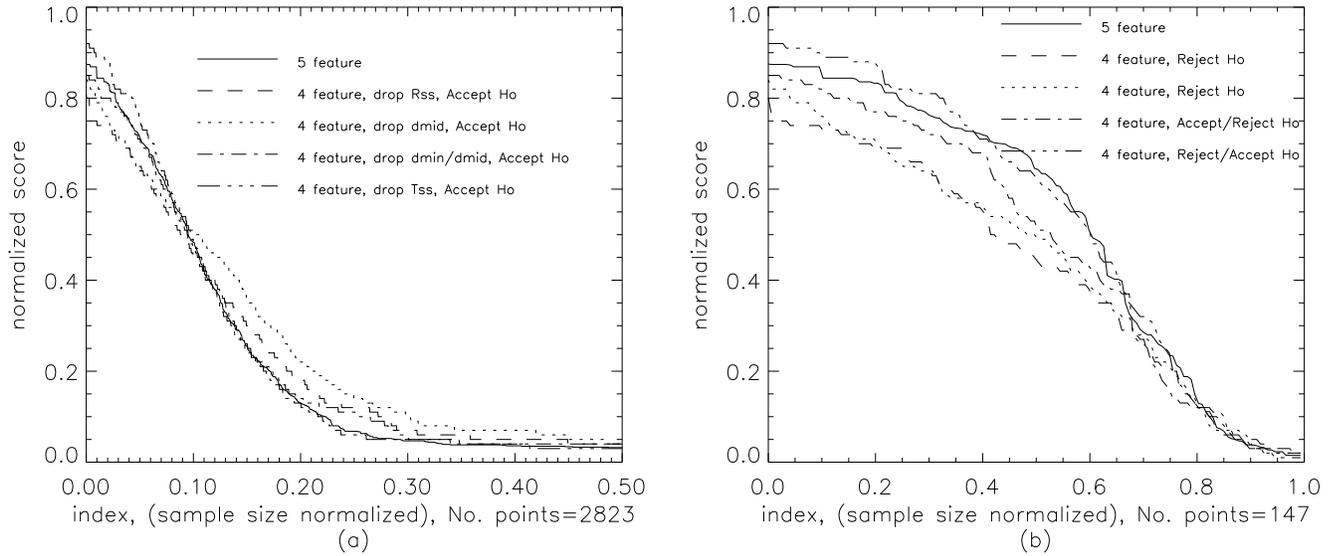}
 \caption{\label{fig:epsartd} Vote curve comparisons of five feature classifier with its various four feature classifier subsets, (a) entiretraining/test set, (b) visual bents.  The excluded feature is listed in (a).  OC1 was not successful in separating classes when ($d_{mid}$+$d_{min}$)/$d_{max}$ was dropped.} \label{Fd}
\end{figure*}

Figure~\ref{Fd} shows the comparison of the five feature set with its various four member feature subsets.  The feature being dropped is indicated in the legend in part (a) of the figure.  OC1 was not
successful in separating classes when the bentness ratio, ($d_{mid}$+$d_{min}$)/$d_{max}$,
was dropped from the five feature set.  The significance test results show the training/test set vote curves for the successful four feature classifiers are not significantly different from the five feature classifier at the 5\% level.  As shown in Figure~\ref{Fd}b, dropping $R_{ss}$ and $d_{mid}$ resulted in significantly different and degraded visual bent vote curves, indicating necessity of these members of the feature set, whereas dropping $d_{min}$/$d_{mid}$ and $T_{ss}$ showed mixed results.  Significance tests comparing these latter two four-feature classifiers directly did show significant differences.  Thus, $T_{ss}$ is chosen for deletion, since that showed the most compact distribution for the training/test set and generally higher scores for the visual bents.

\clearpage

\subsection{\it Comparison of Four-Member Feature Set with its Various Three-Member Feature Subsets}
\begin{figure*}[h]
 \centering
 \includegraphics{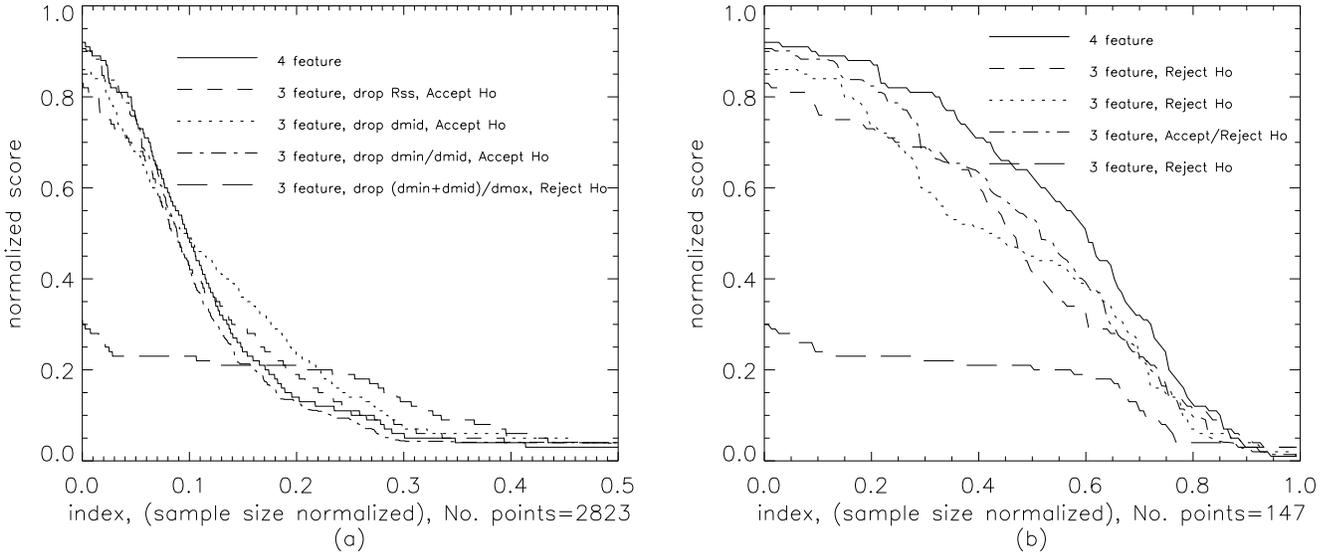}
 \caption{\label{fig:epsarte} Vote curve comparisons of four feature classifier with its various three feature classifier subsets, (a) entire training/test set, (b) visual bents.} \label{Fe}
\end{figure*}

To examine even simpler feature sets, decision trees were
attempted dropping, in succession, each of the four features of previous best four feature set.  A comparison of the vote distributions are shown in Figure~\ref{Fe}.  Again, the legend in part (a) of the figure indicates the dropped feature.  
Dropping the projected arm length ratio, $d_{min}$/$d_{mid}$ appears to have the least effect on the training/test set vote curve, whereas dropping the bentness ratio, ($d_{mid}$+$d_{min}$)/$d_{max}$, has the most deleterious effect. Dropping $R_{ss}$ and $d_{mid}$ have intermediate effects.  Significance test results are as shown.  As for the four feature classifiers, features $d_{mid}$, ($d_{mid}$+$d_{min}$)/$d_{max}$ and $R_{ss}$ are needed, with $d_{min}$/$d_{mid}$ of perhaps more marginal necessity.  
In order to explore three feature comparisons, $d_{min}$/$d_{mid}$ will be dropped as a feature.

\clearpage

\subsection{\it Alternative Forms for Variables}

At this point it is of interest to compare classifications resulting from the best three feature set ($d_{mid}$, ($d_{mid}$+$d_{min}$)/$d_{max}$, $R_{ss}$) with an eight feature set ($d_{mid}$, ($d_{mid}$+$d_{min}$)/$d_{max}$, the six constituent catalog variables of $R_{ss}$).   The constituent catalog variables being the three used in equation (1) for each of the two non-core components.
This comparison examines the ability of the classifier to deal with complex functional relationships.

\begin{figure*}[h]
 \centering
 \includegraphics{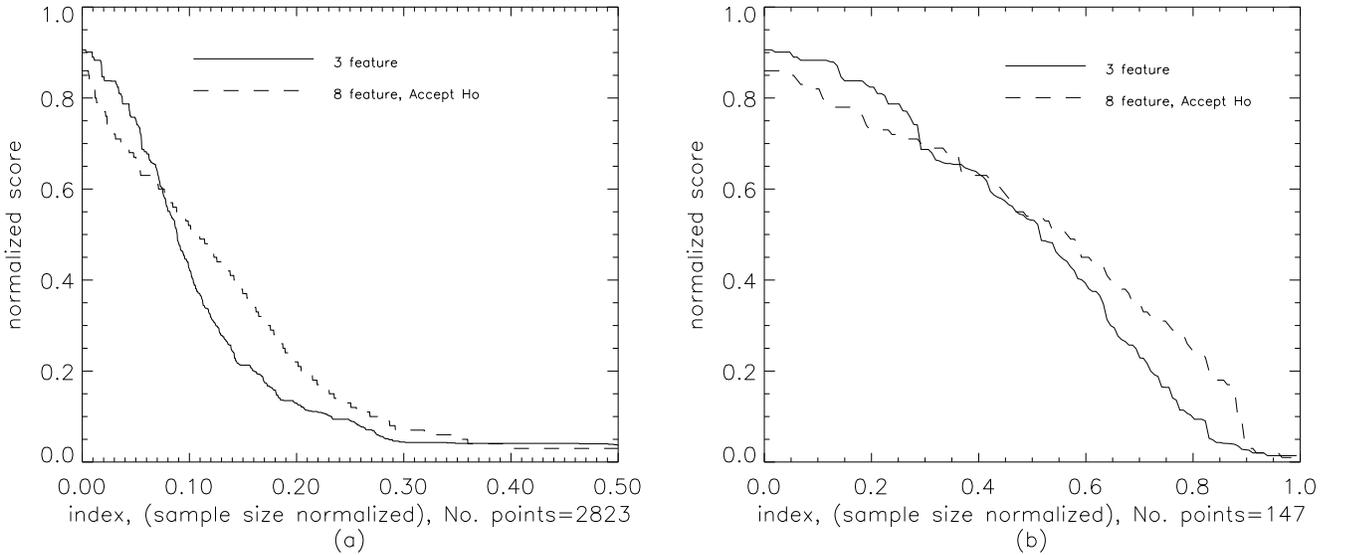}
 \caption{\label{fig:epsartg} Vote curve comparisons of three feature classifier with expanded-form eight feature classifier, one of three features expanded in terms of its six components, (a) entire training/test set, (b) visual bents.} \label{Fg}
\end{figure*}

 The vote curves for this feature set comparison are shown in Figure~\ref{Fg}.  As might be expected, the three feature training/test set distribution appears more compact, though it is not significantly different at the 5\% significance level. 
The hypothesis tests show at the 5\% level, the visual bent distributions are equivalent.  
This seems a rather powerful example of the ability of the decision tree classifier to adapt to
different functional forms of the features, assuming all relevant information is available.
There is again considerable scatter in the direct vote comparison for the visual bent doubles (not shown).

\begin{figure*}[h]
 \centering
 \includegraphics{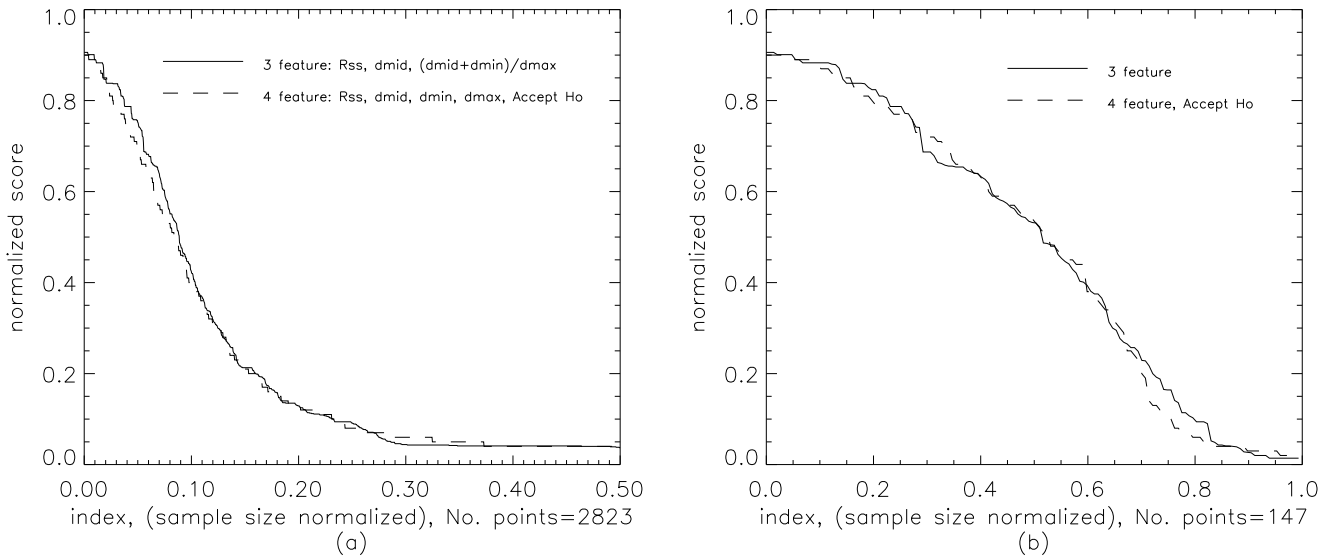}
 \caption{\label{fig:epsartf} Vote curve comparisons of three feature classifier with four feature expanded-form classifier, (a) entire training/test set, (b) visual bents.  } \label{Ff}
\end{figure*}

A second alternative-forms comparison is for the three feature set ($R_{ss}$, $d_{mid}$, ($d_{mid}$+$d_{min}$)/$d_{max}$)) compared with the four feature set ($R_{ss}$, $d_{min}$, $d_{mid}$, $d_{max}$).  These comparisons are shown in Figure~\ref{Ff}. 
Here, the visual bent vote curve is nearly identical for the two forms and the scatter is somewhat reduced in the direct vote comparison (not shown). 

\subsection{\it Classifier Generation Comparison} 
\begin{figure*}[h]
 \centering
 \includegraphics{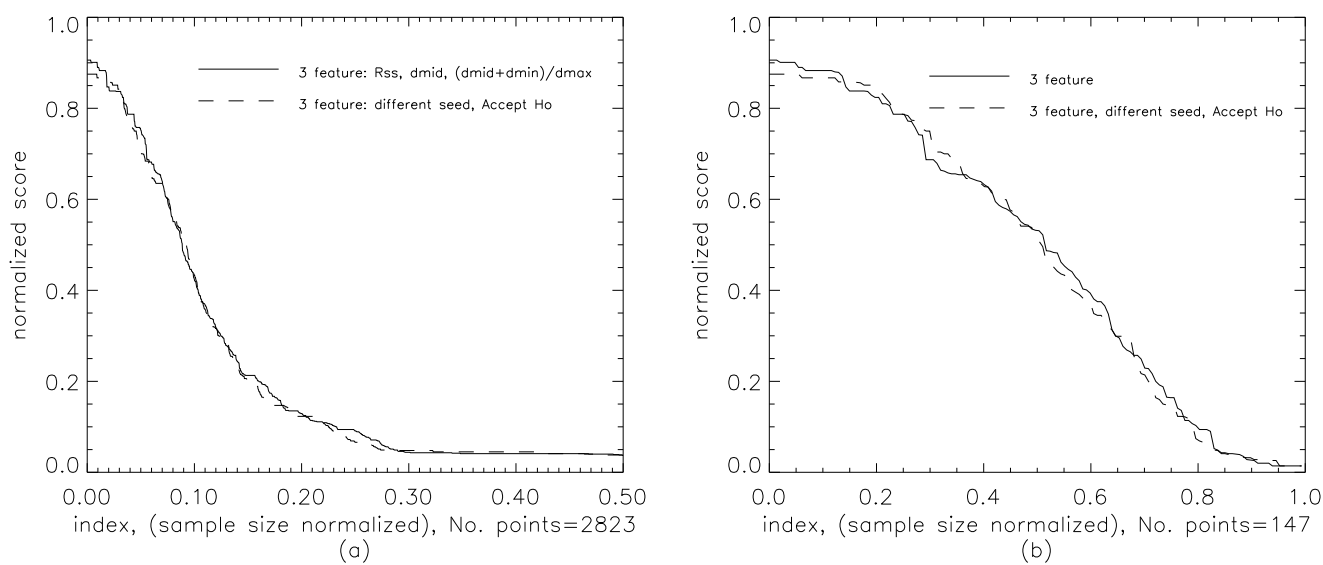}
 \caption{\label{fig:epsartk} Vote curve comparisons of two separate generations of three feature classifier, (a) entire training /test set, (b) visual bents.} \label{Fk}
\end{figure*}

In order to examine the sensitivity of the vote to decision tree generation, a separate five-fold, ten-classifiers-per-fold, decision tree ensemble was generated using different random number seeds for the above best three feature classifier.  The vote curves are shown in Figure~\ref{Fk}.  Note the continuous interweaving of the distributions in Figure~\ref{Fk}, in contrast to previous comparisons.  

The use of Tsao's truncated Smirnov's test for the training/test sets will now be discussed.  Initial hypothesis tests using Kolmogorov-Smirnov and Wilcoxon signed rank tests on curves in Figure~\ref{Fk}a resulted in rejection of the hypothesis of equivalent distributions, clearly not the expected or desired result.  This rejection appears to be an artifact of the relatively small number of folds and the quantization of decision tree results, there being large numbers of a few small but slightly different values for the two generations.  Since details of the vote curves below say 0.05 are not of particular interest, the curves above that value were compared using Tsao's truncated Smirnov's distribution \citep{BTs} as developed by \citet{BCo}.  A random sample of 60 points from each training/test set was examined.  Using this statistic, the hypothesis of equivalent visual vote curves is accepted.  It is this significance test result that is reported in Figure~\ref{Fk}a.  

\begin{figure}[h]
 \centering
 \includegraphics{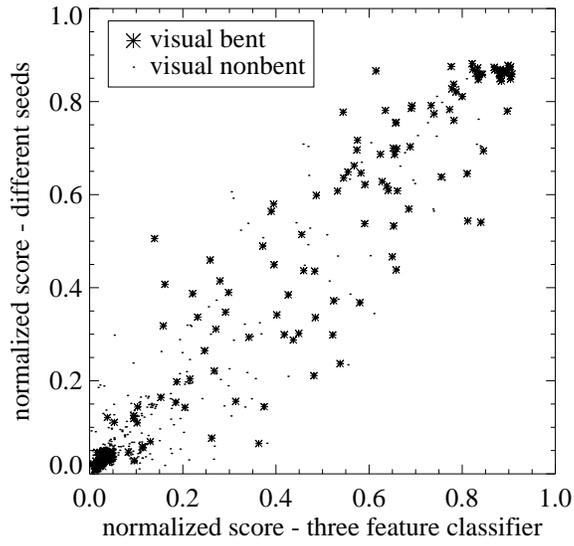}
 \caption{\label{fig:epsartl}Vote comparison of two separate generations of three feature classifier.} \label{Fl}
\end{figure}

In Figure~\ref{Fl}, the direct vote comparison,  the higher vote sources show better agreement than previous cases, suggesting classifier generation using five folds with ten classifiers per fold is a less significant source of error than the feature set selection.  Direct vote comparison with 20 initializations per fold, five fold classifiers  and 10 initializations per fold, 20 fold classifiers showed similar scatter, suggesting feature set selection or visual classification a larger source of error than classifier generation.  Examination of the scatter in the classifications of a training set of half size showed similar variation to the full training set, again suggesting visual classification and inadequacy of feature set the largest source of error.  Comparison of the vote curves for half-size training set classifiers with full-size training set classifiers showed non-significant differences at the 5\% level. 

\clearpage

\subsection{\it Two Member Feature Sets}
\begin{figure*}[h]
 \centering
 \includegraphics{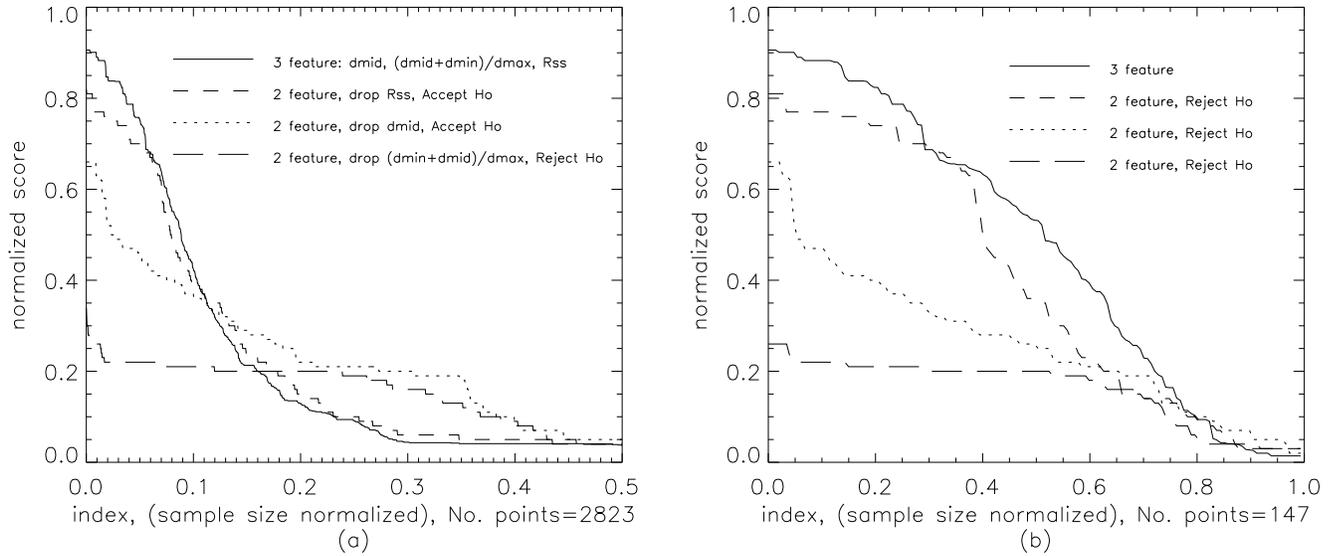}
 \caption{\label{fig:epsarth}Vote curve comparisons of three feature classifier with its various two feature classifier subsets, (a) entire training/test set and (b) visual bents.} \label{Fh}
\end{figure*}

Next, the various two feature subsets of the above best
three feature set are compared in Figure~\ref{Fh}.  Again
dropping the bentness ratio, ($d_{mid}$+$d_{min}$)/$d_{max}$, has the most significant
impact.  
The significance tests on the visual bent vote curves reiterate the necessity for all three features.

\clearpage

\subsection{\it Feature Space Plots and Result Comparison}

\begin{figure*}[!h]
 \centering
 \plotone{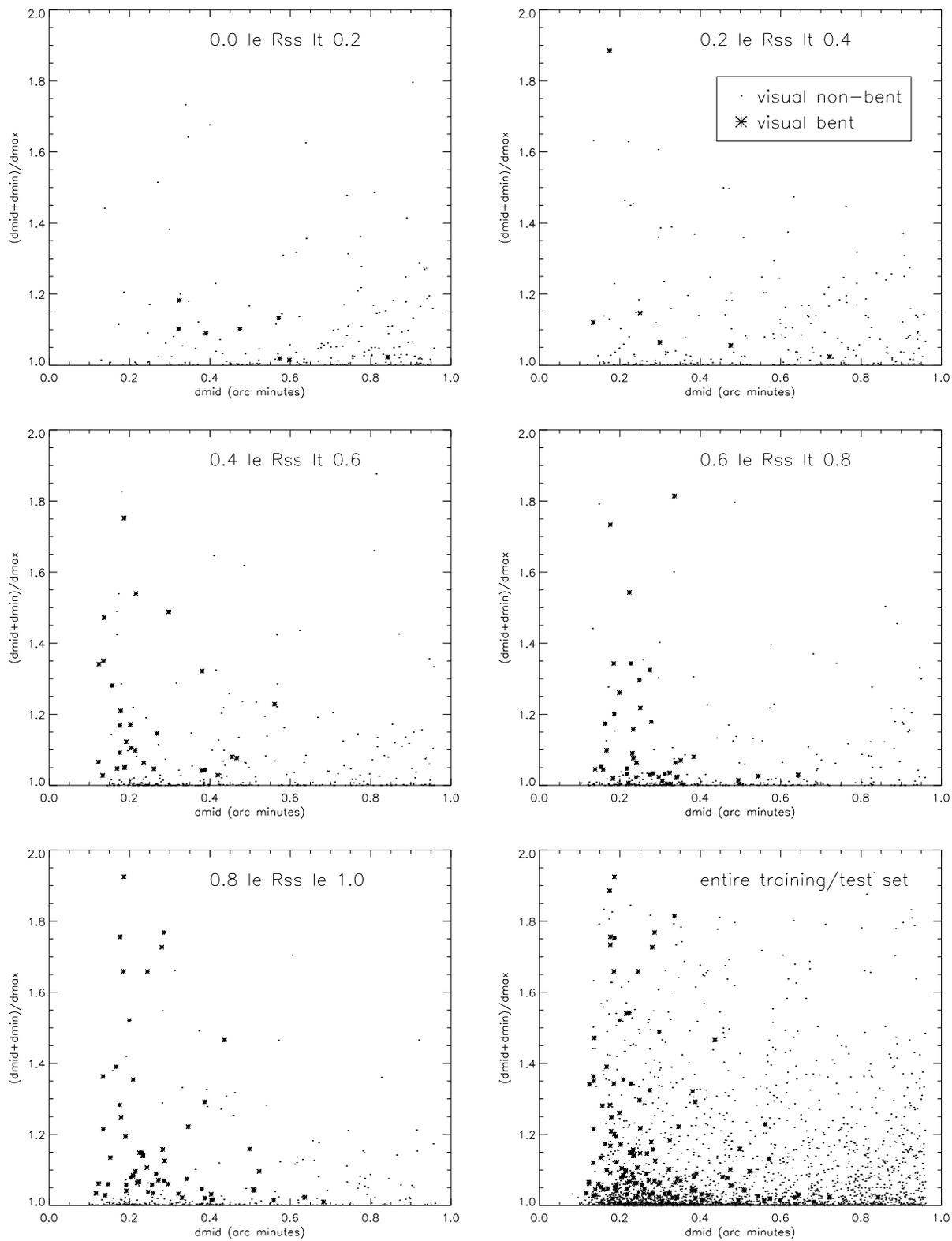}
\caption{\label{fig:epsarti}Visual bent and nonbent sources as function of $d_{mid}$ and ($d_{mid}$+$d_{min}$)/$d_{max}$ for various $R_{ss}$ ranges.} \label{Fi}
\end{figure*}

\begin{figure*}[!h]
 \centering
 \plotone{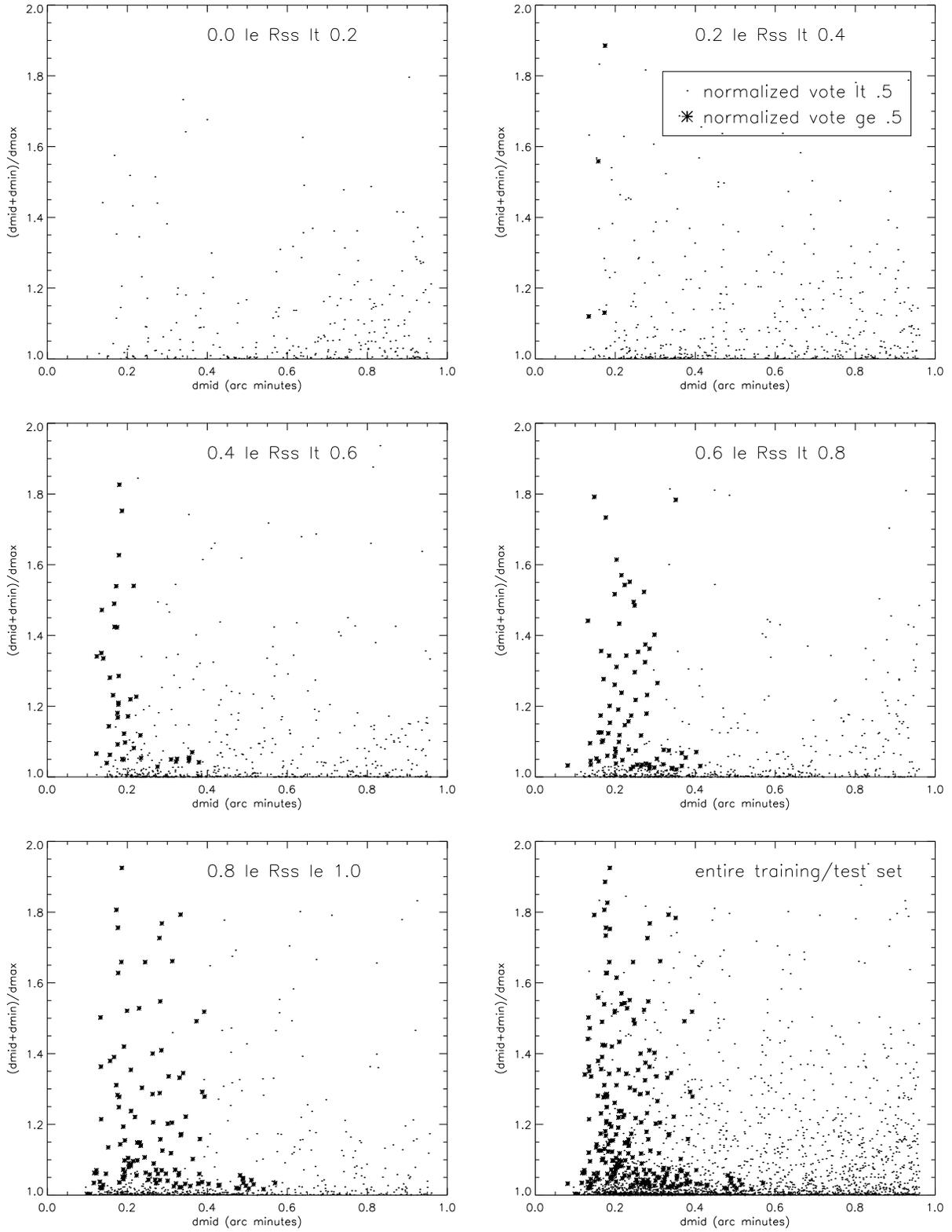}
 \caption{\label{fig:epsartj}Vote comparison as function of $d_{mid}$ and ($d_{mid}$+$d_{min}$)/$d_{max}$ for various $R_{ss}$ ranges.} \label{Fj}
\end{figure*}

For the above best three feature classifier, as an alternative to detailed examination of the fifty decision trees in the ensemble, two dimensional visualization can be employed to deduce the region of feature space occupied by the target class.
  Figure~\ref{Fi} and 
Figure~\ref{Fj}
show plots of $d_{mid}$ vs. ($d_{mid}$+$d_{min}$)/$d_{max}$ for various $R_{ss}$ intervals.
Figure~\ref{Fi} shows the visual bent and nonbent classifications, while
Figure~\ref{Fj} shows sources with vote greater than 0.5 as bold.
Overall results are as might be expected, in that the target class has higher bentness ratio and ratio of silhouette sizes closer to one.
However, best boundary values would have been difficult to determine without
pattern recognition algorithms.  It is noted that re-examination of sources classified as bent in the two top plots of Figure~\ref{Fi} suggest they may be some of the more dubious visual classifications.  

\clearpage

\clearpage

\begin{figure*}[!h]
 \centering
  \plotone{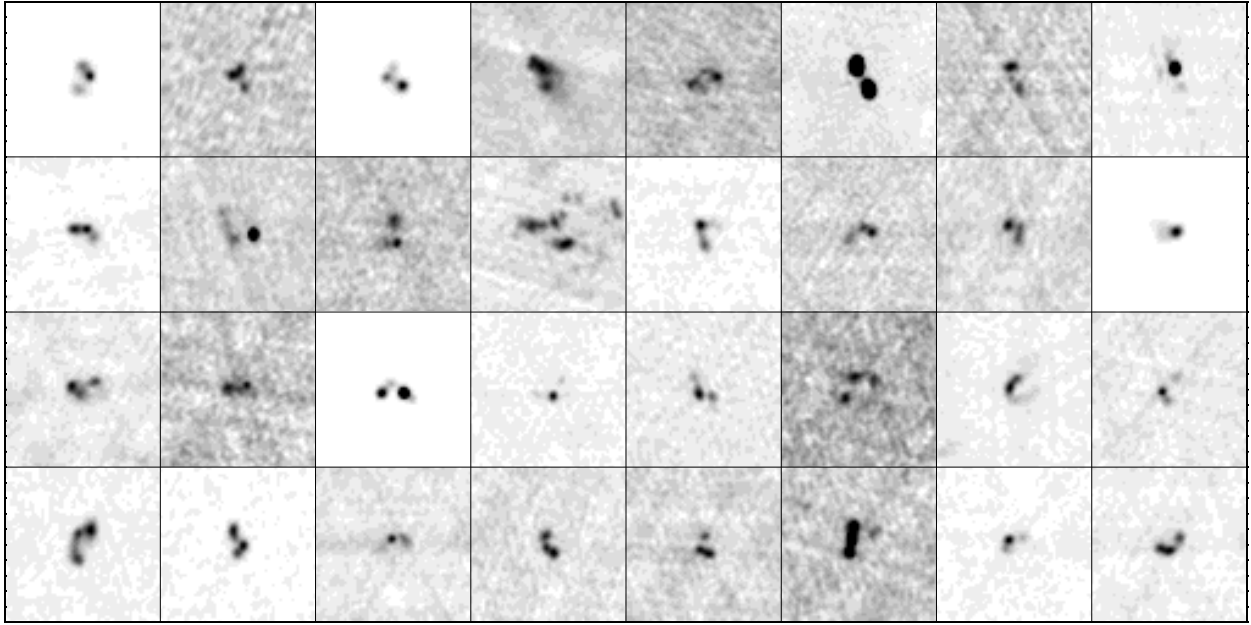}
  \caption{\label{fig:epsarts}A random selection of 32 highest ranked sources in entire catalog, best four feature classifier, training sources excluded.} \label{Fs}
\end{figure*}

\begin{figure*}[!h]
  \centering
  \plotone{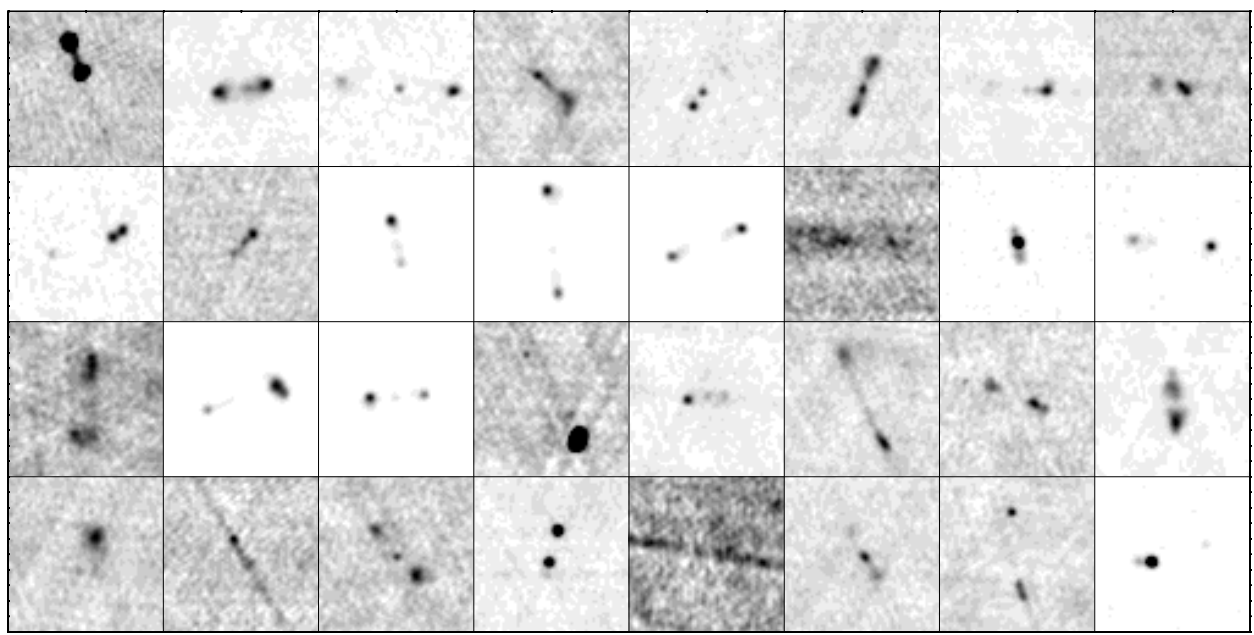}
  \caption{\label{fig:epsartt}Random selection of 32 of lowest ranked sources in entire catalog, best four feature classifier, training sources excluded.} \label{Ft}
\end{figure*}

Finally Figure~\ref{Fs} shows  32 sources randomly selected  from highest ranked sources (vote value =0.86) of the best four feature classifier applied to the entire catalog available at the time.  No training set sources are included.  These can be compared with Figure~\ref{Ft} showing 32 randomly selected lowest ranked sources (vote value =0.03) from that classifier.  Results seem consistent with respective estimated probabilities.

\clearpage

\section{Discussion and Conclusions}

Specific feature set comparisons have been demonstrated using the sort-ordered, sample-size-normalized vote distribution of an ensemble of decision trees in conjunction with the visual bent vote curves.  While recognition rates and classification errors may be adequate for feature set comparison in some applications, the vote curves provide a method of comparison for applications where they are problematical.  Indeed, they should also be useful in cases where accurate classifications are available but feature sets are inadequate.

Though failure in classifier construction was observed with the exclusion of an apparently essential feature, significant degradation of results due to redundant or irrelevant features was not found for this application and training set.
Adding as many as sixteen features to the original set did not have a major effect.  
One instance was observed where dropping a feature resulted in somewhat improved compactness of the vote distribution.  Dropping the total silhouette size, $T_{ss}$, from the five feature set, demonstrated marginal improvement with deletion of a feature.

Though for each alternative forms comparison, the lower-count feature set produced the more compact vote curve, the results were not significantly different at the 5\% level.  Using the six component variables of the silhouette size ratio as a features  in place of the ratio seems a powerful demonstration of the ability of the decision tree classifiers to deal with complex functional relationships.

Vote curve analysis provides a method to evaluate the effect of training set size, number of folds and number of classifiers per fold on classification errors.  Using multiple classifiers per fold allows error estimation on the probability of a sample being of the target class, given the training set, classifier, and feature set.  While OC1 was the particular decision tree system used in this study, the method would be applicable to other systems employing randomization in generation of the classifiers.

Of the feature sets examined, the four feature set, {$d_{mid}$, ($d_{mid}$+$d_{min}$)/$d_{max}$, $R_{ss}$}, {$d_{min}$/{$d_{mid}$, provided the most desirable visual bent vote distribution, though the {$d_{min}$/{$d_{mid}$ feature is of arguable necessity.  While it is easy to understand the incorporation of the latter three features, the necessity for inclusion of $d_{mid}$, a feature that in some sense sets a scale, is more interesting.  However, since an observationally verified training set is unavailable, it cannot be ruled out that this is a selection effect.

It is noted the optimal feature subset may not have been found.
It is expected that would require a generally infeasible exhaustive search or, if applicable, use of
branch and bound \citep{BNF} techniques.  However, as a practical matter,  the best three and four feature classifiers presented here significantly reduce the number of sources to be examined in the search for bent doubles. 


\acknowledgments
Thanks are due the anonymous referee for the suggestions for improved clarity and questions that lead to the tying up of several loose ends.

R. Becker provided computer resources.  Richard White provided software to access the FIRST images as well as discussion of vote apportionment for pruned decision trees.  Extensive use was made of the IDL astro package (http://idlastro.gsfc.nasa.gov).  The free availability of the OC1 decision tree software was greatly appreciated (anonymous ftp from ftp.cs.jhu.edu directory pub/oc1).
  The author is grateful for office space and computing facilities provided by the Institute of Geophysics and Planetary Physics (IGPP), John Bradley, director, and Kem Cook.  The term 'vote curve' was coined by an anonymous referee of a previous paper.

This work was performed under the auspices of the U.S. Department of Energy,
National Nuclear Security Administration by the University of California, Lawrence Livermore National Laboratory under contract No. W-7405-Eng-48.



\clearpage

\end{document}